# Rectangle-like hysteresis in a Dysprosium Metallacrown Magnet with Linear F–Dy–F Anisotropic Moiety

Si-Guo Wu,[1,∇] Ze-Yu Ruan,[1,∇] Jie-Yu Zheng,[1] Guo-Zhang Huang,[1] Veacheslav Vieru,[2,4] Yan-Cong Chen,[1] Le Tuan Anh Ho,[2,3] Jun-Liang Liu,[1] Liviu F. Chibotaru,*[2] and Ming-Liang Tong*[1]

[1]Key Laboratory of Bioinorganic and Synthetic Chemistry of Ministry of Education, School of Chemistry, Sun Yat-Sen University, 510275 Guangzhou, Guangdong, P. R. China.

[2]Theory of Nanomaterials Group, Katholieke Universiteit Leuven, Celestijnenlaan 200F, 3001 Leuven, Belgium.

[3]Department of Chemistry, National University of Singapore, 3 Science Drive 3 Singapore 117543.

[4]Faculty of Science and Engineering, Maastricht University, Lenculenstraat 14, 6211 KR Maastricht, The Netherlands.

**ABSTRACT:** Single-molecule magnets (SMMs) exhibiting open hysteresis loops may potentially apply to molecule-based information processing and storage. However, the capacity to retain magnetic memory is always limited by zero-field quantum tunneling of magnetization (QTM). Herein, a well-designed dysprosium metallacrown SMM, consisting of an endohedral approximate linear F–Dy–F strong anisotropic moiety in a peripheral [15-MC$_{Ni}$-5] metallacrown (MC), is reported with the largest reversal barrier of 1060 cm$^{-1}$ among d-f SMMs. Rectangle-like hysteresis loops are observed with the huge squareness (remanence/saturation magnetization) up to 97% at 2 K. More importantly, zero-field QTM step is phenomenologically removed by minimizing the dipole coupling and hyperfine interactions. The results demonstrate for the first time that zero-field QTM step can be eliminated via manipulating the ligand field and vanishing the external magnetic perturbations, which illuminates a promising blueprint for developing high-performance SMMs.

Single-molecule magnets (SMMs), being classified among nanomagnets, have drawn great attention to scientists for their potential applications in molecular electronic and magnetic devices.[1] In recent years, great strides were achieved since the discovery of several performant lanthanide based-SMMs with large effective energy barriers ($U_{eff}$) and correspondingly high blocking temperatures ($T_B$).[2] However, there remains a formidable challenge against the zero-field quantum tunneling of magnetization (QTM) which is featured as a sudden collapse or even vanishment of remanence in hysteresis.[3] The conservation of zero-field QTM inevitably restrains the magnetic memory effect therefore hinders the practical application in data storage.[4] Thus, sufficiently inhibition of zero-field QTM is an imperative demand for designing high-performance SMMs.

The QTM event, essentially originated from transverse crystal field and/or external magnetic perturbations, is characterized by demagnetization in SMMs since the tunneling gap surpasses the Zeeman splitting.[5] Transverse crystal field allow the mixing of $m_J$ sublevels and consequently facilitate QTM effect.[6] Additionally, external magnetic perturbations, including dipole[5,7] and hyperfine interactions[8], make zero-field QTM more noticeable especially for Kramers ions. For one thing, the relaxation dynamics of individual molecule is affected by the dipolar field deriving from adjacent spins. The distribution of dipolar field alters upon the surrounding spins reversal, which makes more spins into resonance.[9] On the other hand, the interplay between electron and nuclear spins results in the concomitant hyperfine structures. Such hyperfine interactions in Kramers ions destroy the degenerate bistable states and thus promote the tunneling rate.[10] In general, crystallographic restrictions and external magnetic perturbations have significant influences on magnetic relaxation in low-temperature region wherein QTM-dominated process tends to appear. In order to cut down the zero-field QTM step, a judicious strategy needs not only to enforce the lanthanide-ion axial anisotropy, but also to minimize the dipole and hyperfine interactions as possible.

According to the crystal field (CF) theory, QTM should have been completely suppressed by placing a lanthanide ion into specific symmetries of ligand fields (such as $C_{\infty}$ and $D_{5h}$) thanks for the extinction of transverse CFs.[6,11] Taking advantage of metallacrown (MC) approach as well as axially anisotropic ligation[12,13], herein we report a 3d-4f magnet [DyNi$_5$(quinha)$_5$F$_2$(dfpy)$_{10}$](ClO$_4$)·2EtOH (**1**, H$_2$quinha = quinaldichydroxamic acid, dfpy = 3,5-difluoropyridine). In complex **1**, the F–Dy–F moiety en-cages into a weak but rigid equatorial ligand field provided by the [15–MC$_{Ni}$–5]. The MC ring provides an appropriate environment for dysprosium ion with $D_{5h}$ local symmetry. Meanwhile, 10 dfpy ligands, located on both sides of MC ring, prevent 2 "naked" F$^-$ anions from bridging or generating hydrogen bonds. DyF$_2$[15-MC$_{Ni}$-5] retains a Kramers system as a whole, which theoretically ensures at least doubly degeneracy of magnetic states and the absence of zero-field QTM. In this work, complex **1** was characterized by single-crystal X-ray diffraction, SQUID magnetometry and *ab initio* calculations. Moreover, to evaluate the contribution of dipolar/hyperfine coupling toward zero-field QTM, yttrium diamagnetic diluted analogue {Dy$_{0.06}$Y$_{0.94}$Ni$_5$} (**2**) and the further isotopically pure sample {$^{164}$Dy$_{0.05}$Y$_{0.95}$Ni$_5$} (**3**, $I$ = 0) were investigated through isothermal magnetization as well as time-dependent decay of magnetization measurements.

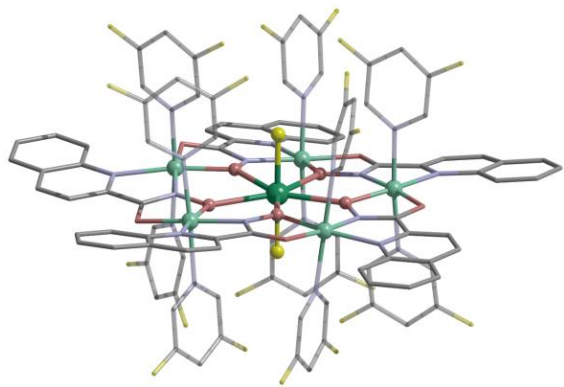

**Figure 1.** The molecular structure of **1**. The anion, solvent molecules and hydrogen atoms are omitted for clarity. Colour code: dark green (Dy), light green (Ni), grey (C), red (O), blue (N) and yellow (F).

The structure of **1** is similar to its holmium analogue.[12] In complex **1** (Figure 1), the tetradentate ligand quinha$^{2-}$ form five-membered chelate rings with Ni$^{II}$ ions in a "head-to-tail" way, giving rise to a neutral [15–MC$_{Ni}$–5]. 3,5-difluoropyridine ligands coordinate with Ni$^{II}$ ions on both sides of MC ring, which results in a hydrophobic cavity. As in such arrangement, the Dy$^{III}$ ion is equatorially captured within the cavum by 5 hydroximate oxygen atoms from the MC ring while axially capped by 2 terminal fluoride ions. In this case, the Dy$^{III}$ possesses a compressed pseudo pentagonal bipyramid coordination environment with average Dy–O and Dy–F bond lengths of 2.455 Å and 2.128 Å, respectively. In addition, the equatorial O–Dy–O angles lie in the range of 71.13(6)–72.47(6)° and the axial F–Dy–F angle is 176.77(7)°, in accord with the nearly perfect $D_{5h}$ geometry verified by CShM calculation[14] (Table S3). All the Ni$^{II}$ ions are 6-coordinate with a [N$_4$O$_2$] environment featuring octahedral geometry. Molecules are well-separated with the shortest intermolecular Dy…Dy distance of *ca.* 13.11 Å.

Direct-current (dc) magnetic susceptibility measurements were performed on polycrystalline sample under 1000 Oe dc field. The thermal susceptibility product $\chi_M T$ value is 19.68 cm$^3$ K mol$^{-1}$ at 300 K (Figure S5), close to the sum of one free Dy$^{III}$ ion (14.17 cm$^3$ K mol$^{-1}$, $^6H_{15/2}$, $g_J$= 4/3) and the experimental $\chi_M T$ of its yttrium analogue.[12] As the temperature lowers, the $\chi_M T$ value gradually decreases to *ca.* 13 cm$^3$ K mol$^{-1}$ until 8 K, due to the splitting of $^6H_{15/2}$ ground state as well as the antiferromagnetic interactions among Ni$^{II}$ ions (see Supporting Information for details). After that, a rapid drop is found at lower temperature attributed to magnetic blocking, which can be further confirmed by the divergence of zero-field cooled/field cooled (ZFC–FC) magnetization (Figure S7).

The temperature dependence of the alternative-current (ac) out-of-phase susceptibility ($\chi''_M$) exhibits peaks with distinct temperature and frequency dependency between 0.1–1488 Hz (Figure 2 and Figure S8–9). The maximum peak of $\chi''_M$ lies at 87 K and no "tails" are observed at low temperatures even for 0.1 Hz, suggesting zero-field QTM is indeed significantly precluded. Notably, field-dependency of magnetization dynamics at 26 K (Figure S12) reveals that complex **1** shows the slowest relaxation rate at zero dc field, confirming the zero-field QTM is indeed minimized. In other word, complex **1** is an ideal candidate for information storage device as it maintains the longest relaxation time in the absence of an applied field. The Arrhenius plot of magnetic relaxation times shows two linear regions among 30–91 K (Figure 2b).

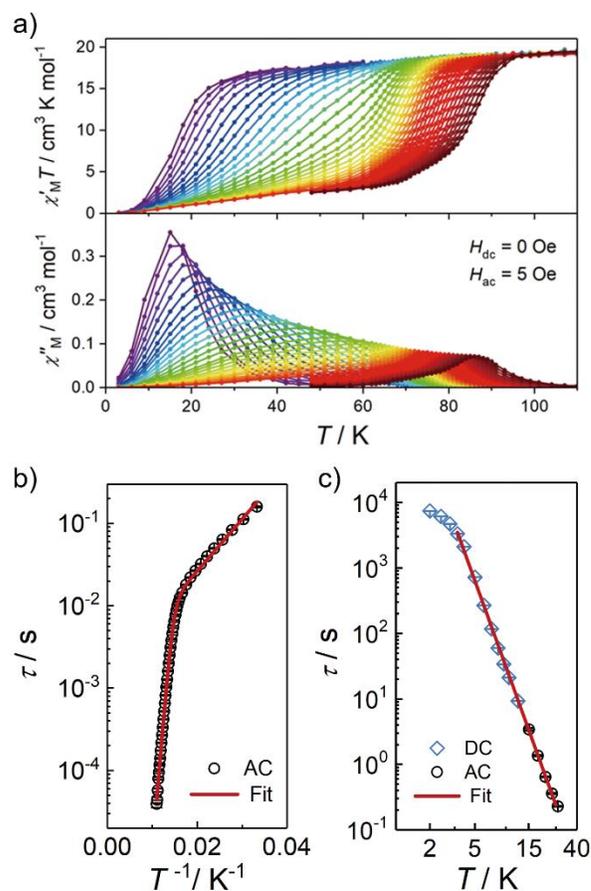

**Figure 2.** a) The temperature dependence of the in-phase product ($\chi'_M T$) and out-of-phase ($\chi''_M$) for **1** at zero dc field with the ac frequency of 0.1–1488 Hz. b) Arrhenius plot of relaxation time in zero dc field at 30–91 K. c) Temperature

dependence of the relaxation time in zero dc field for **1** at 2–27 K. The red solid line at high temperature domain is the best fit by using a sum of two Orbach processes, whilst Raman process was used for the low temperature region.

Arrhenius fits by using equation $\tau^{-1} = \tau_{o(1)}^{-1}\exp(-U_{eff(1)}/k_B T) + \tau_{o(2)}^{-1}\exp(-U_{eff(2)}/k_B T)$ to give $U_{eff(1)} = 1060(11)$ cm$^{-1}$, $\tau_{o(1)} = 3.7(7) \times 10^{-12}$ s, $U_{eff(2)} = 101(2)$ cm$^{-1}$ and $\tau_{o(2)} = 1.5(1) \times 10^{-3}$ s. To the best of our knowledge, $U_{eff}$ of 1060 cm$^{-1}$ is much higher than most reported SMMs and yields a new record for d-f SMMs, far surpassing the mark of 577 cm$^{-1}$ set by its holmium analogue.[12] In comparison to the recently reported complexes [Dy(Tp$^{py}$)F(dioxane/py)](PF$_6$)[15] and [Dy(L)F](CF$_3$SO$_3$)$_2\cdot$H$_2$O[16], F–Dy–F moiety shows much stronger axial anisotropy than Dy–F unit with more than doubled reversal barrier (436 cm$^{-1}$/79 cm$^{-1}$). More importantly, complex **1** exhibits rectangular hysteresis loops thanks to the well-designed ligand field (*vide infra*), whereas waist-restricted ones for [Dy(Tp$^{py}$)F(dioxane/py)](PF$_6$) as a result of significant zero-field QTM.

Further investigations toward magnetic relaxation mechanisms of **1** were obtained through *ab initio* calculations (see Supporting Information for details). The complete active space self-consistent field spin-orbit (CASSCF-SO) calculations confirm the strong anisotropy of F–Dy–F moiety since the ground and the first excited doublets is well separated by 419 cm$^{-1}$. The relaxation pathway follows the sequence of $m_J$= ±15/2, ±13/2, ±11/2, ±9/2 and ±7/2 and gives the magnetization blocking barrier of ~943 cm$^{-1}$, slightly lower than the experimentally observed one (Figure S33). Thus, the high-temperature Orbach process arises from the single-ion behavior of individual dysprosium. For low temperature domain, the exponential $\tau(T)$ regime may be assigned to the spin-phonon modulation involving the low-lying exchange Kramers doublets.[12]

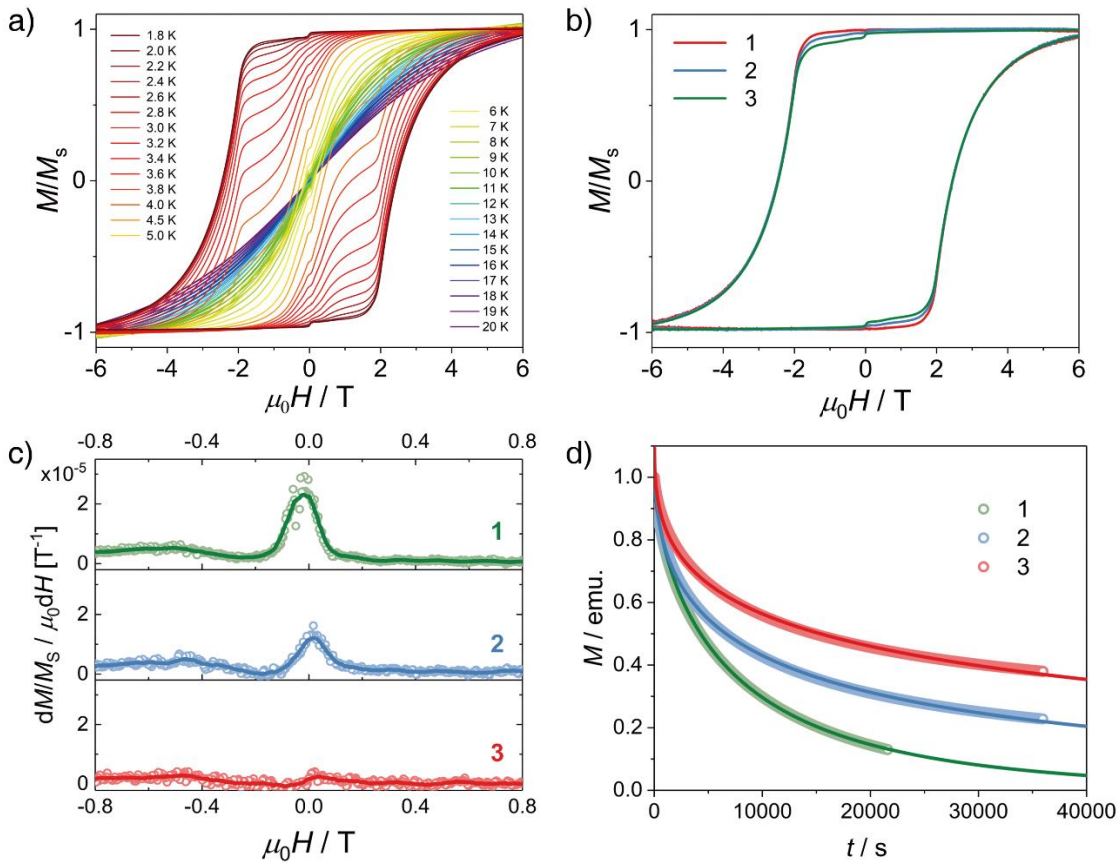

**Figure 3.** a) Normalized hysteresis for **1** using an average sweep rate of 0.02 T s$^{-1}$. b) Isothermal magnetization at 2 K for (**1**) DyNi$_5$, (**2**) Dy$_{0.06}$Y$_{0.94}$Ni$_5$ and (**3**) $^{164}$Dy$_{0.05}$Y$_{0.95}$Ni$_5$ with a sweep rate of 0.02 T s$^{-1}$. c) The corresponding 1$^{st}$ differential of hysteresis at 2 K. d) The data of time-dependent decay of magnetization at 2 K, the solid lines represent for the best fit using the stretched exponential function.

To gain further insight into low-temperature magnetic relaxation process, we determined the $\tau$ for 2–14 K via time-dependent decay of magnetization. The relaxation times, fitted by the stretched exponential function, increase drastically with decreasing temperature and reach 7452(1) s at 2 K, meaning that the spin reversal is frozen intensely (Figure S19–21 and Table S8). Arranging the $\tau$ below 30 K on a log-log scale gains a near-linear correlation plot (Figure 2c), referring to a Raman-dominated process which meets the power law ($\tau^{-1} = CT^n$) with $C$ = 7.1(7) × 10$^{-7}$ s$^{-1}$ K$^{-4.8}$ and $n$ = 4.8(1). At lower temperature (< 3 K), $\tau$ becomes temperature independent since some faster magnetic relaxation processes, such as QTM or phonon bottleneck effect[17], starts to appear.

Isothermal magnetization for **1** was recorded ranging from –6 T to 6 T with a sweep rate of 0.02 T s$^{-1}$. Surprisingly, the hysteresis loop at 1.8 K shows a large coercivity of 2.46 T and a squareness up to 97% (Figure 3a), indicating QTM is efficiently suppressed at zero field. As the zero-field step is intensely involved with sweeping rate, variable-sweeping-rate hysteresis loops were carried out (Figure S15), revealing that the remanence ($M$r) retains ~91% even at a sweeping rate as slow as 0.001 T s$^{-1}$. Noted that the squareness outperforms most reported SMMs, in stark contrast to those high-barrier SMMs with small or even loss of remanence. Rectangular hysteresis with large $M$r would be of great significance in practical applications for the magnetic memory effects at zero dc field.[18] In addition, the hysteresis loops remain open till 20 K, manifesting a new record for d-f SMMs far surpassing that was held by {Cr$_2$Dy$_2$}[19] (4.7 K, Table S7). Amid slowly increasing the temperature, a clear platform forms at zero field position. Differential of hysteresis between 1.8 K–5 K (Figure S16–17) displays a positive peak at nearly zero field < 3 K whereas a negative peak > 3 K. Videlicet, fast zero-field QTM is sufficiently suppressed until 3 K, which matches well with the aforementioned relaxation mechanism in relevant temperature region.

According to our previous study[12], the ground state of {Ni$_5$} ring is diamagnetic due to the zero-spin projection. Therefore, zero-field QTM at extremely low temperature might be ascribed to the external magnetic perturbations. To examine the corresponding contributions from dipolar or hyperfine interactions, we prepared the ~5% diluted sample (**2**) and the isotopically enriched one (**3**, $^{164}$Dy, 96.0% purity) for contrast, and recorded the hysteresis loops (2 K ,0.02 T s$^{-1}$) as well as the $\tau$ value by dc relaxation measurements. As showed in Figure 3b, complexes **1–3** reveal similar magnetic hysteresis except the remanence which enhances to 98.2% for **2** and 99.6% for **3**. The changes of 1$^{st}$ differential curves (Figure 3c) also clearly indicated the inhibition of dipole and hyperfine interactions. The remanence in **3** is slightly off hundred percent, which might mainly come from the 4% natural abundance components and/or imperfect $D_{5h}$ coordination geometry of Dy$^{III}$ site. Moreover, the magnetization decay curves (2 K, zero dc field, Figure 3d) demonstrate a smaller relaxation rate for **2** and **3**, the relaxation time increases to 11264(39) s and 26474(127) s, respectively. Notably, it is the first example to nearly remove zero-field QTM step in SMMs by minimizing the dipole coupling and hyperfine interactions. The results from both hysteresis and dc relaxation confirm the non-negligible impact from external magnetic perturbations.

To summarize, DyF$_2$[15-MC$_{Ni}$-5], bearing the F–Dy–F unit, was obtained by rational design of ligand field. Thanks for the massive crystal-field splitting engendered by the difluoride ligation, a record-high barrier among d-f SMMs reaches 1060 cm$^{-1}$. Magnetization measurements reveal a rectangle-like hysteresis at low temperature due to efficient suppression of QTM event. More importantly, the tiny zero-field QTM step is further minimized via $^{164}$Dy enrichment in a 1:19 diamagnetic matrix. The results observed here highlight that not only dipolar coupling but also hyperfine interactions are the dominant causes for QTM regime. In addition, this work provides a new perspective for designing high-performance SMMs by the combination of an adequate axial anisotropy and the minimization of external magnetic perturbations.

## AUTHOR INFORMATION

### Corresponding Author

**Ming-Liang Tong** − *Key Laboratory of Bioinorganic and Synthetic Chemistry of Ministry of Education, School of Chemistry, Sun Yat-Sen University, 510275 Guangzhou, Guangdong, P. R. China;* orcid.org/ 0000-0003-4725-0798; Email: tongml@mail.sysu.edu.cn

**Liviu F. Chibotaru** − *Theory of Nanomaterials Group, Katholieke Universiteit Leuven, Celestijnenlaan 200F, 3001 Leuven, Belgium;* Email: Liviu.Chibotaru@kuleuven.be

### Authors

**Si-Guo Wu** − *Key Laboratory of Bioinorganic and Synthetic Chemistry of Ministry of Education, School of Chemistry, Sun*

Yat-Sen University, 510275 Guangzhou, Guangdong, P. R. China; orcid.org/ 0000-0002-0110-8350

**Ze-Yu Ruan** − *Key Laboratory of Bioinorganic and Synthetic Chemistry of Ministry of Education, School of Chemistry, Sun Yat-Sen University, 510275 Guangzhou, Guangdong, P. R. China*

**Guo-Zhang Huang** − *Key Laboratory of Bioinorganic and Synthetic Chemistry of Ministry of Education, School of Chemistry, Sun Yat-Sen University, 510275 Guangzhou, Guangdong, P. R. China;* orcid.org/ 0000-0001-9167-0090

**Jie-Yu Zheng** − *Key Laboratory of Bioinorganic and Synthetic Chemistry of Ministry of Education, School of Chemistry, Sun Yat-Sen University, 510275 Guangzhou, Guangdong, P. R. China*

**Veacheslav Vieru** − *Theory of Nanomaterials Group, Katholieke Universiteit Leuven, Celestijnenlaan 200F, 3001 Leuven, Belgium; Faculty of Science and Engineering, Maastricht University, Lenculenstraat 14, 6211 KR Maastricht, The Netherlands;* orcid.org/ 0000-0001-6375-4537

**Yan-Cong Chen** − *Key Laboratory of Bioinorganic and Synthetic Chemistry of Ministry of Education, School of Chemistry, Sun Yat-Sen University, 510275 Guangzhou, Guangdong, P. R. China;* orcid.org/ 0000-0001-5047-3445

**Le Tuan Anh Ho** − *Theory of Nanomaterials Group, Katholieke Universiteit Leuven, Celestijnenlaan 200F, 3001 Leuven, Belgium; Department of Chemistry, National University of Singapore, 3 Science Drive 3 Singapore 117543;* orcid.org/ 0000-0002-3284-502X

**Liviu F. Chibotaru** − *Theory of Nanomaterials Group, Katholieke Universiteit Leuven, Celestijnenlaan 200F, 3001 Leuven, Belgium;* orcid.org/ 0000-0003-1556-0812

Author Contributions

$^{\nabla}$These authors contributed equally to this work.
Notes
The authors declare no competing financial interest.

ACKNOWLEDGMENT

This work was supported by the National Key Research and Development Program of China (2018YFA0306001), the NSFC (Grant nos 21620102002, 21822508, and 21821003), and the Pearl River Talent Plan of Guangdong (2017BT01C161).

SYNOPSIS TOC

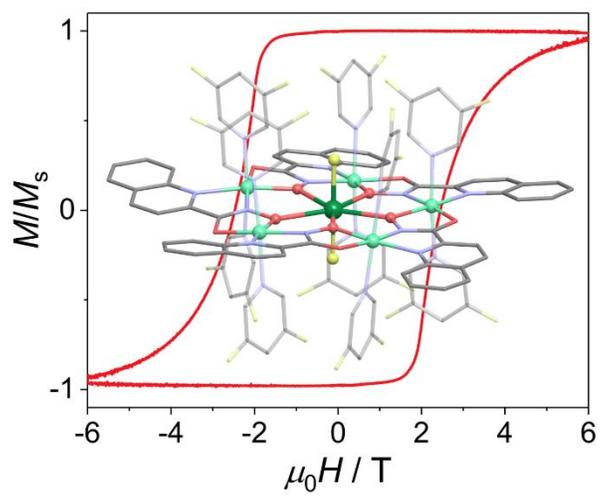